# Integrating AI and Learning Analytics for Data-Driven Pedagogical Decisions and Personalized Interventions in Education

**Ramteja Sajja**[1,3] (ramteja-sajja@uiowa.edu)
**Yusuf Sermet**[3] ( msermet@uiowa.edu)
**David Cwiertny**[2,3,4,5] ( david-cwiertny@uiowa.edu)
**Ibrahim Demir**[1,2,3] (ibrahim-demir@uiowa.edu)

[1] Department of Electrical Computer Engineering, University of Iowa
[2] Department of Civil and Environmental Engineering, University of Iowa
[3] IIHR Hydroscience and Engineering, University of Iowa
[4] Department of Chemistry, University of Iowa
[5] Center for Health Effects of Environmental Contamination, University of Iowa

**Abstract**
This research study explores the conceptualization, development, and deployment of an innovative learning analytics tool, leveraging OpenAI's GPT-4 model to quantify student engagement, map learning progression, and evaluate diverse instructional strategies within an educational context. By analyzing critical data points such as students' stress levels, curiosity, confusion, agitation, topic preferences, and study methods, the tool provides a comprehensive view of the learning environment. It also employs Bloom's taxonomy to assess cognitive development based on student inquiries. In addition to technical evaluation through synthetic data, feedback from a survey of teaching faculty at the University of Iowa was collected to gauge perceived benefits and challenges. Faculty recognized the tool's potential to enhance instructional decision-making through real-time insights but expressed concerns about data security and the accuracy of AI-generated insights. The study outlines the design, implementation, and evaluation of the tool, highlighting its contributions to educational outcomes, practical integration within learning management systems, and future refinements needed to address privacy and accuracy concerns. This research underscores AI's role in shaping personalized, data-driven education.

## 1. Introduction

In the rapidly transforming realm of education, the confluence of learning analytics (LA) and artificial intelligence (AI) signifies a critical paradigm shift. The transformative capabilities of these technologies and their potential to revolutionize the educational landscape spur new research (Ouhaichi et al., 2023). Interdisciplinary research drawn from learning sciences provides an invaluable foundation to probe and comprehend human learning processes. The amalgamation of this knowledge with AI technologies holds the potential to considerably enhance pedagogical methodologies (Luckin and Cukurova, 2019). Recent advancements in AI, particularly the development of large language models like GPT-4, have showcased potential applications in various educational contexts (Pursnani et. al., 2023). These models have been utilized to improve performance in standard examinations, highlighting the importance of noninvasive prompt modifications and revealing the remarkable progress in their mathematical capabilities. Such advancements pave the way for AI's potential in solving complex engineering and health problems (Sermet and Demir, 2018; 2021), and educational challenges. AI's evolving role in education is exemplified by the development of tailored AI-enabled educational assistants that integrate with Learning Management Systems (LMS) to provide personalized learning experiences, especially in fields requiring complex data interactions (Sajja et. al., 2024). Such advancements highlight the growing impact of AI in enhancing student engagement and comprehension in specialized domains like environmental sciences.

However, a gap persists as many AI developers exhibit limited familiarity with learning sciences research, highlighting an imperative need to foster collaborative relationships among AI developers, educators, and researchers. Moreover, the strength of Natural Language Processing (NLP) as an augmentative tool for LA is gaining recognition. While LA has made considerable strides, the incorporation of qualitative, textual data through NLP could introduce a richer, contextual understanding, adding depth and nuance to our grasp of learning processes (Smith et. al., 2022).

With the steady integration of AI within the education sector (Sermet and Demir, 2020), it is crucial to comprehend its impact and application. Alam (2022) offers a particularly insightful exploration of AI's evolution from computer technology to intelligent online education and embedded computer systems. AI not only enhances the efficiency and quality of instruction but also tailors individual learning pathways by adapting curriculum to meet each student's unique needs. In this context, Ouyang et. al. (2023) provides a noteworthy examination of AI performance prediction models within education. The study identifies a gap where existing AI models tend to prioritize algorithm development and optimization, often overlooking the need for timely, continuous feedback to students. The collective insights from these studies underscore the motivation behind our research - to harness the potential of AI within LA to optimize educational outcomes.

Our study aims to achieve several primary objectives. Firstly, we aim to develop an AI-empowered LA tool designed to effectively measure and analyze student engagement. This technology's goal is to monitor and quantify various aspects of student involvement, ultimately enabling the delivery of pedagogical interventions based on real-time, actionable insights. Secondly, we seek to construct a comprehensive AI-based system to monitor students' learning progression. This objective relies on harnessing AI's predictive modeling capabilities to map individual learning pathways, providing educators with a nuanced understanding of each student's educational journey. Lastly, we aim to establish a mechanism to assess students' affective states. By leveraging AI capabilities, we aim to gain insights into the emotional context of learning, recognizing its significant role in shaping cognitive processes and educational outcomes.

Our research boundaries are explicitly outlined, with a central focus on the integration of AI into LA for improved educational outcomes. In addressing the challenges of educational equity, our tool is designed with a focus on inclusivity, ensuring that insights derived from AI analytics are accessible and beneficial to diverse student populations. The specific elements under consideration include the measurement of student engagement, tracking of learning progression, and assessment of affective states. Evaluated metrics comprise, but are not limited to, participation rates, progression speed, depth of understanding, emotional response, and learning adaptation.

The structure of this paper is as follows: Section 2 delves into the existing literature, emphasizing gaps in the current knowledge. Section 3 articulates the methodology, detailing the design choices and implementation of the LA tool. Section 4 presents our findings and interprets their significance. Finally, Section 5 discusses the strengths and limitations of our study, potential future directions, and summarizes our contributions to the field.

## 2.   Related Work

Learning analytics, an emerging interdisciplinary field, stands at the crossroads of various sophisticated domains including machine learning, data science, education, cognitive psychology, among others. Initiated with the intent to utilize the vast data produced through technological integration in education, LA has grown into a community with substantial impacts on research, practice, policy, and decision-making (Gašević et. al., 2017; Baker & Inventado, 2014). This composite field leans on its theoretical underpinnings, design considerations, and data science methodologies to offer innovative perspectives for enhancing learning (Reimann, 2016; Siemens, 2013). Despite being in its growth phase, LA has demonstrated potential for transforming teaching practices, informing learning research, and influencing the educational landscape (Clow, 2013; Dawson et. al., 2019).

The expansion of LA is stimulated by an explosion of accessible learner data and concurrent management methodologies emphasizing quantitative metrics. This data availability allows for a more nuanced comprehension of student behavior and performance, which can aid educators in efficient resource allocation, thereby augmenting teaching outcomes (Clow, D.). LA encompasses a plethora of data types, each requiring unique analytical techniques and methodologies. The application spectrum ranges from eye-tracking to automated online dialog analysis, from ecosystem surveys to log data analysis at individual and collaborative levels, and even extends to visual LA applied to IoT data (Nistor & Hernández-García, 2018).

Recently, deep learning models have made significant contributions to knowledge tracing, a vital component of LA. Deep Knowledge Tracing (DKT) uses data to forecast student performance, facilitating early identification of potential learning hurdles and enabling precise intervention strategies (Casalinoa et. al., 2021). The field deploys various methods such as visual data analysis techniques, social network analysis, and educational data mining, all aiding in targeted course offerings, curriculum development, personalized learning experiences, and improving instructor performance. Yet, challenges persist in data tracking, evaluation, and analysis, along with ethical and privacy concerns, and the need for a stronger integration with learning sciences to optimize learning environments (Avella et. al., 2016).

In recent studies, there has been a concerted focus on student engagement. Both traditional and AI-enhanced educational environments are being meticulously analyzed. Bowden et. al. (2021) suggest a holistic approach to dissect student engagement, examining involvement and expectations as antecedents to engagement and gauging the impact of engagement on various student and institutional outcomes. Conversely, Goldberg et al. (2021) employ machine learning to assess visible engagement during classroom instruction. They substantiate a manual rating system, demonstrating the potential efficacy of a machine vision-based approach using gaze, head pose, and facial expressions as engagement indicators. In the realm of e-learning, Bhardwaj et al. (2021) utilize deep learning algorithms to track students' real-time emotional responses. By integrating facial landmark detection, emotional recognition, and survey results, they compute a Mean Engagement Score (MES), enhancing the effectiveness of digital learning methodologies. Lastly, Nkomo et. al. (2021) presents a comprehensive systematic review of the impact of digital technologies on teaching and learning practices over the past decade, highlighting notable gaps in our understanding and measurement of student engagement with digital technologies.

Despite notable advancements in learning analytics (LA), several gaps remain. Current LA methods often lack integration with pedagogical theory and teaching practice, making it challenging for educators to apply these insights effectively. Moreover, LA faces issues with scalability and generalizability across different educational settings. Additionally, many LA tools do not adequately address the affective aspects of learning, which are crucial for understanding student engagement.

The proposed AI-enhanced LA tool could bridge these gaps. Informed by pedagogical theory, this adaptable tool could be sensitive to emotional states, mindful of ethical and privacy considerations, and capable of offering actionable insights to educators. By incorporating advanced AI and machine learning techniques, the assistant could analyze and respond to student data in real-time, providing immediate, personalized feedback and support for students and invaluable insights for educators.

## 3.    Methodology

This research centers around the development of a novel LA tool, conceptualized as an extension to existing educational platforms such as educational intelligent assistants. The tool's primary purpose is to enhance the capacity of these platforms to collect, process, and analyze a wealth of data, extending beyond the information traditionally gathered by Learning Management Systems (LMS). VirtualTA, an AI-augmented intelligent educational assistant, plays a pivotal role in this study. VirtualTA is specifically designed to address logistical questions about courses (Sajja et al., 2023a) and provide responses to course-content related inquiries. Additionally, it can generate flashcards and quizzes (Sajja et al., 2023b).

While the development and application of the tool proposed in this study is contextualized within the VirtualTA environment, it is designed with flexibility and adaptability in mind, making it compatible with a multitude of educational chatbots, and smart assistants such as Instant Expert (Sermet and Demir, 2019) or LMS. The overarching aim of the tool is to offer educators a comprehensive understanding of students' performance, engagement, and learning patterns. This, in turn, enables them to make data-driven decisions to enrich the overall learning experience. The research proceeds systematically, encompassing stages of data collection, processing, analysis, and finally, tool development and deployment.

## 3.1.    Data Collection

The data collection method for the LA tool is designed to gather extensive information from interactions with an educational chatbot, like VirtualTA, and is adaptable to other chatbots or Learning Management Systems (LMS). It captures all student interactions, including raw dialogues and an emotional analysis of each input. This analysis assesses the affective states in student communications, offering insights into the emotional aspects of their learning experience.

The tool also records the topics of questions and messages, helping map academic interests and identify areas of confusion or difficulty. The VirtualTA is equipped with features like flashcard creation and quizzes, and the method closely monitors user engagement with these aspects. It tracks variables such as time spent on questions, overall duration on quizzes, and quiz outcomes, providing data on students' study preferences, subject understanding, and engagement with different learning methods.

The strategy combines quantitative data (like student metadata, assignment scores, quiz results, time on learning materials, and interaction frequency from the LMS) with any available qualitative data (such as instructor-conducted surveys). This comprehensive approach yields a holistic view of student behavior, preferences, and attitudes towards their learning process.

### 3.1.1.    Data Pre-processing and Analysis

Before delving into detailed analysis, it's essential to recognize that raw data, despite its depth and breadth, requires rigorous preprocessing. This step is fundamental to address any inconsistencies and rectify potential inaccuracies. Such careful preparation entails a series of critical measures designed to bolster the reliability of the data. The data collection process begins when a user interacts with the chatbot. Once the session concludes, either through an exit or quit action, the accumulated data is forwarded to the backend in JSON format. In this phase, the primary focus is to analyze user queries, evaluating affective states and learning progression.

A significant part of the initial processing is data cleaning. This involves pinpointing and rectifying or removing irrelevant, duplicated, or incorrect data entries. This step is not just about quality enhancement; it's a protective measure against any potential misinterpretations or inaccuracies in future analyses. For instance, irrelevant data such as instances where a "user clicked on a button" may not offer any meaningful insights. Similarly, incorrect data entries, like instances of starting a quiz but not finishing it, can lead to inaccurate time measurements and provide no clarity on how many questions were answered correctly or incorrectly. Post-analysis, the system determines whether to append the student's data to an existing record or initiate a new entry if one doesn't already exist. The refined data is then updated on the instructor's dashboard, reflecting the most recent metrics. This preprocessing ensures that the insights presented to the instructors are both accurate and derived from a clean, consistent dataset.

### 3.1.2. *Ethical Data Recording and Privacy Considerations*

In the LA tool's development, ethical considerations, particularly regarding student data privacy, are prioritized. The tool records time-series data to track student engagement and performance over time, but it does not store the actual content of student messages to protect privacy and foster a safe environment for help-seeking. It focuses on analyzing interactions to compute metrics like emotional states, recording data points only during active interactions. Once these metrics are computed, the raw text data is discarded to maintain privacy.

By concentrating on computed metrics rather than raw text, the tool not only respects the confidentiality of student communications but also aligns with 'privacy by design' principles. This ensures that any future applications involving real data will have the necessary infrastructure for data protection and student anonymity from the outset. Moreover, the temporal collection strategy allows the tool to serve as a dynamic gauge of the students' academic and emotional journey, enabling educators to monitor trends and identify critical periods that may require intervention. This data, while anonymized and aggregated to protect individual identities, provides educators with valuable insights to tailor their teaching strategies and better support their students.

## 3.2. Metrics and Indicators

In student behavior analysis, a range of metrics and indicators are employed including student engagement, affective emotional states, and learning progression. This comprehensive suite of metrics (Table A1) allows us to take a holistic view of student interactions and experiences, with each metric providing a unique perspective on their learning journey. Using advanced emotional analysis techniques, these emotional states are inferred from the collected data. These inferred emotional states, however, are not definitive. They act as indicators, providing valuable insights into each student's learning experience, including areas of interest, struggles, and progress.

Emphasis should be placed on engagement and learning progression. Engagement metrics assist with understanding the degree of student participation and interaction, thereby indicating their level of interest and commitment. On the other hand, learning progression metrics provide insights into academic improvement over time, giving a measure of the effectiveness of the learning experience. However, it's essential to emphasize that these metrics and indicators are not diagnostic tools and should not be used to categorize students. They offer valuable insights but should always be interpreted with a full understanding of their limitations.

To ensure clarity and transparency, the tool conveys the following disclaimer where appropriate: "*This data is not 100% factual and should be used as a reference only.*"

### 3.3.    Development of Learning Analytics Tool

The design and development of the LA tool leverages the power of OpenAI's GPT-4 (OpenAI, 2023) to enhance the learning experience. This tool seamlessly integrates with existing virtual teaching assistants and other educational intelligent assistants. The integration with VirtualTA serves as one of the use cases, highlighting the tool's compatibility with different educational platforms. The core functionality of this LA tool extends beyond the VirtualTA, offering a wide array of analytical capabilities across different domains.

One of the key features includes sentiment analysis on students' questions collected from the VirtualTA interface. Leveraging the power of the GPT-4 model, specially configured for few-shot learning, and enhanced with prompt engineering techniques, our tool assesses the emotional undertones in these questions and scores are given on a scale of 1 to 10 for stress, agitation, curiosity, and confusion. Prompt engineering customizes the model's response to specific types of questions, enhancing its sensitivity and accuracy in recognizing subtle emotional cues. This offers a real-time assessment of the emotional atmosphere in the learning environment, enabling rapid identification and intervention for areas of high confusion or stress.

In addition to sentiment analysis, the model identifies the topic related to each question, providing valuable insights into the subjects that students find particularly challenging or engaging. This could help educators identify trends in student inquiries, thus informing more effective curriculum design and resource allocation. The tool also tracks the progression of learning by analyzing the questions using Bloom's taxonomy once the students exit the VirtualTA interface. This allows for the assessment of cognitive development and guides personalized learning experiences.

Moreover, the tool integrates with the LMS to gain access to a wider range of data points. Information such as student metadata, assignment scores, quiz results, time spent on different learning materials, as well as usage data including access times, quiz duration, quiz topic, time spent on questions, and students' preferred study methods, is incorporated. Thus, the tool offers a comprehensive, responsive, and dynamic analytics system, illustrating an innovative application of GPT-4 in the field of education.

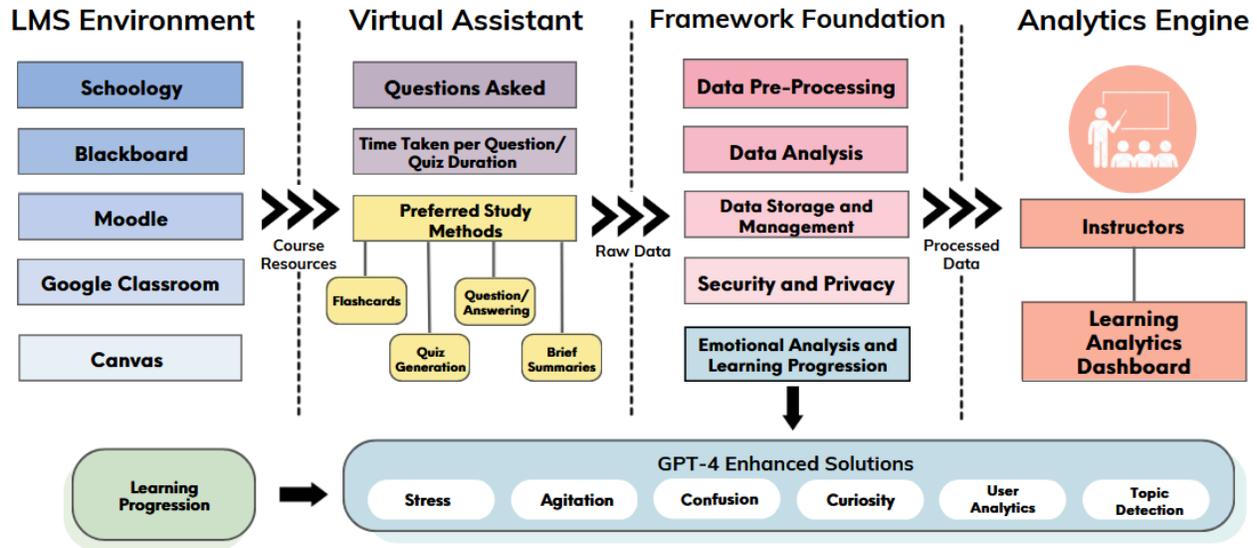

**Figure 1**. System Architecture Diagram and Components

As portrayed in Figure 1, the LA tool seamlessly integrates with the LMS employed by educators, incorporating a Virtual Assistant to extract course resources from the LMS environment. This system is designed to meticulously gather raw data, which is then channeled through a series of stages—pre-processing, analysis, and secure storage—within the LA tool. To enrich this data further, a Smart AI integration is implemented, undertaking an emotional analysis, and learning progression guided by Bloom's Taxonomy. Following these operations, the refined data is encapsulated into a structured JSON format and sent to the front-end application. This mechanism empowers instructors with the capability to visualize and comprehend the data, consequently fostering data-driven pedagogical advancements.

### 3.4. Implementation and Deployment

The presented LA tool integrates with various educational platforms, analyzing students' interactions to generate insights. This adaptability ensures compatibility with educational chatbots and LMS. The primary focus during the tool's development and deployment was to transform interactions into data points for meaningful analysis.

### 3.4.1. Natural Language Inference

The LA tool has harnessed the capabilities of the GPT-4 model to delve into the nuanced tones embedded within user queries, enabling the assessment of stress levels, agitation, curiosity, and confusion. GPT-4 represents a significant advancement in the field of natural language processing, distinguished by its remarkable natural language understanding capabilities and its aptitude to discern broader contextual intricacies, surpassing its predecessors in these regards. This heightened contextual comprehension holds a pivotal role in the tool's operation.

To derive these essential metrics, we have employed few-shot learning with a temperature setting of 1.0. Few-shot learning facilitates the rapid adaptation of the model across diverse contexts without the need for extensive fine-tuning. On the other hand, prompt engineering refines the model's interactions, ensuring that it responds more accurately to the specific requirements of the educational domain. The temperature setting of 1.0 ensures that the model's predictions are diverse, allowing for a broader range of responses while still maintaining relevance. By configuring GPT-4 in this manner and harnessing the power of prompt engineering, we equip the LA tool with the capacity to understand the emotional nuances encapsulated within student queries, encompassing stress levels, agitation, curiosity, learning progression and confusion. This approach not only ensures the versatility of our tool but also underscores its ability to provide meaningful insights across various educational scenarios.

In terms of the conversation history, the proposed tool is intended to consider all prior communication between students and the educational intelligent assistant. However, given GPT-4's token limitation — with a maximum of 8192 tokens, equating to approximately 6000 words — we've adopted a first-in-first-out approach. This strategy ensures that the most recent messages remain within the conversation history, offering a snapshot of the latest interactions. By leveraging this conversation history, we can trace the learning progression of students, grounded in the established framework of Bloom's taxonomy. The conversation history illustrates the tool's capability to manage and process data.

### 3.4.2.  *Software Development and Architecture*

The LA dashboard has been designed using vanilla JavaScript, ensuring a user interface that's both straightforward and efficient for instructors. To aid in visualizing student data patterns and offer a more interactive experience, we've integrated Chart.js. This library is known for its capability to produce dynamic, real-time charts that assist instructors in easily discerning trends and insights. The technical foundation of our system relies on the synergistic combination of Express.js and Node.js in the backend. This duo not only streamlines backend operations but also optimizes performance, ensuring that the system remains responsive even with high data loads.

One of the essential tasks of our backend infrastructure is to handle and process data from the educational intelligent assistant, which in this context is provided by VirtualTA, as well as to interface directly with OpenAI's GPT API. Communication with these educational assistants is facilitated through APIs of the LA tool, ensuring a seamless transfer of data. All the information sourced from these chatbot interactions is structured in the JSON format. This choice was made considering JSON's inherent flexibility and ease of integration. After thorough processing and analysis, the data is then relayed to the dashboard, ensuring that instructors have timely and easy access to the insights they need.

### 3.5.  Case Study Design

The AI-driven learning analytics tool was evaluated through a survey distributed to teaching faculty at the University of Iowa. The primary aim was to gather feedback on the tool's potential effectiveness in enhancing educational outcomes and its compatibility with existing Learning Management Systems (LMS) such as Canvas. This evaluation focused solely on faculty perceptions of the tool's potential to improve student engagement and provide valuable real-time insights for instructional decision-making. It is important to note that the faculty did not interact directly with the tool, and the feedback gathered was based on their understanding of its described features.

Feedback was collected through an anonymous online survey provided to the participating faculty members. The survey was designed to capture insights across several key areas:

- **Importance of Classroom Insights:** Faculty were asked to select all relevant insights they considered important for understanding student behavior and performance in their classrooms.
- **Student Privacy and Data Security:** Participants shared their concerns regarding student data privacy and security when using new AI-driven tools.
- **Accuracy of Tool-Provided Insights:** Faculty were asked to rate the importance of receiving accurate insights from new tools, ensuring reliability in their teaching practices.
- **Primary Teaching Area and Role:** Instructors indicated their primary teaching area and their role within the institution, to better contextualize their feedback.
- **Additional Comments and Suggestions:** An open-ended question invited faculty to provide any additional comments or suggestions regarding tools designed to improve student learning and engagement.

## 4. Results and Discussion

This section presents the findings from our evaluation of the AI-driven learning analytics tool, focusing on its capabilities and the insights gathered from both synthetic test data and survey responses from teaching faculty at the University of Iowa. The results are structured into two parts: the first evaluates the tool's performance using synthetic data, demonstrating its potential functionalities, while the second presents feedback from a survey of faculty, offering their perspectives on the tool's practical applications and perceived value in educational settings.

We aim to assess how effectively the tool could track student engagement, monitor emotional states, and provide actionable insights for educators. This analysis explores the technical performance of the tool based on synthetic data, alongside faculty perceptions of its potential usefulness and ease of integration into existing pedagogical practices.

### 4.1. Synthetic Data Analysis

The LA tool employs a range of metrics to analyze the multifaceted nature of students' learning experiences. By evaluating emotional states, topic preferences, learning progression, and study methods, the tool aims to provide a holistic perspective of the learning process. Each metric offers insights into different aspects of students' interactions within the learning environment, enabling a nuanced understanding that surpasses traditional assessment methods. To illustrate the tool's capabilities, synthetic test data generated using ChatGPT was utilized. This data, devoid of any actual student interactions, serves to demonstrate the tool's ability to analyze engagement metrics and emotional states while ensuring student privacy and data security.

**Stress, Curiosity, Confusion, and Agitation Measurement:** The LA tool employs GPT-4 for sentiment analysis to evaluate students' emotions during their interactions with VirtualTA, focusing on stress, curiosity, confusion, and agitation. These emotions are scored on a scale of 1 to 10 to reflect intensity. For example, difficulty-related queries may indicate high stress, while exploratory questions suggest curiosity. This real-time emotional analysis helps track students' emotional journeys, highlighting trends and spikes. By capturing these emotional shifts, educators can provide tailored responses, enhancing support and improving the learning environment.

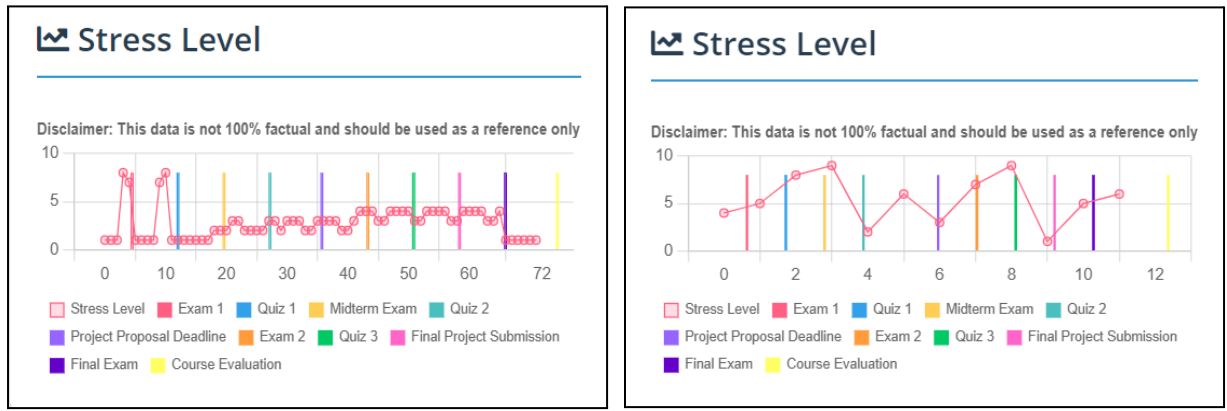

**Figure 2**. Screenshots from the instructor's dashboard (a) Stress Level Graph for overall students, (b) Stress Level Graph for an individual student

The dashboard shows both class-wide (Figure 2a) and individual (Figure 2b) stress levels over time, correlating them with key academic events such as quizzes and assignments. These visualizations help educators understand how emotional states align with the academic schedule, enabling timely interventions to alleviate stress and support student well-being.

**Topic Analysis and Student Questioning Patterns**: The LA tool uses GPT-4 for topic analysis and tracking questioning patterns in student interactions with VirtualTA. It identifies and records the main topics of student queries, mapping out areas of interest or difficulty over time. This enables the identification of patterns, like frequently discussed subjects or challenging topics. Alongside, the tool analyzes questioning patterns, including question types, sentiments, and sequencing. This reveals a student's learning journey, such as progressing from basic to complex questions within a topic, offering insights into their understanding. These analyses allow educators to tailor their teaching and resources more personally, and the system to adapt its responses to each student's learning path.

| ❓ TOPICS ASKED | ❓ TOPICS ASKED |
|---|---|
| Topic: Engineering Ethics | Topic: Ethical Challenges in Sustainability |
| Topic: Impacts of Technological Solutions | Topic: Trust and Credibility of Engineers and Ethical Behavior |
| Topic: Public Health, Safety, and Welfare in Engineering Ethics | Topic: Ethical Codes and Guidelines in Engineering |
| Topic: Confidentiality and Privacy in Engineering Ethics | Topic: Reporting Ethical Violations in Engineering |

**Figure 3**. Topics asked table for overall students in the class.

Figure 3 presents an organized tabular representation of unique topics that students in the class have inquired about up until the point of data collection. For the sake of clarity and efficient use of educators' time, we've ensured that duplicate entries are carefully removed from the table. Each row corresponds to a unique topic, providing educators with a straightforward view of the breadth of subject's students have engaged with throughout the course. This streamlined, tabulated format provides a quick and comprehensive view of the range of topics discussed, aiding them in identifying key areas of student interest or potential difficulties.

**Study Methods and Preferred Learning Approaches:** The LA tool tracks student interactions with VirtualTA, recording activities like asking questions, creating flashcards, taking quizzes, or requesting summaries. This data helps identify individual and class-wide study preferences, such as a preference for active recall through flashcards or learning by testing via quizzes. These insights allow the system to tailor responses and resources to align with each student's habits, enabling a more personalized learning experience. Educators can use this data to adjust their teaching strategies, promoting more effective and diverse study methods.

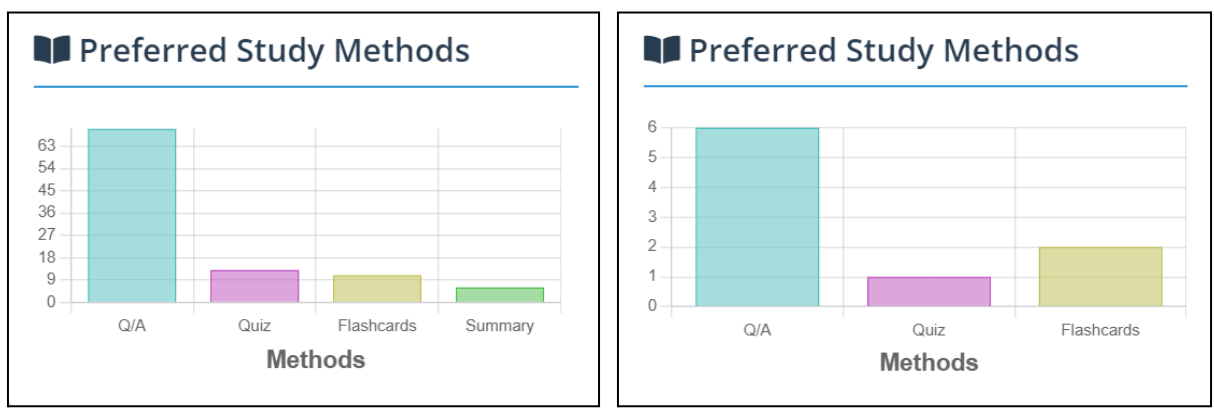

**Figure 4**. Screenshots from the instructor's dashboard (a) Preferred Study Methods for overall students in the class, (b) Preferred Study Methods for a specific student in the class

The graphs show class-wide (Figure 4a) and individual (Figure 4b) study methods, highlighting the most-used interaction modes—questions, quizzes, summaries, and flashcards. This data helps educators understand student preferences, enabling them to tailor instruction and enhance student engagement.

**Tracking and Analyzing Student Interactions**: The LA tool tracks student quiz interactions via VirtualTA, recording quiz topics, completion times, time spent on each question, and quiz results. This data helps analyze students' interests, time management, comprehension speed, and areas of difficulty. The tool allows for personalized learning by adjusting quizzes to match individual strengths and improvement areas. Quiz results also provide insights into student understanding and topic mastery.

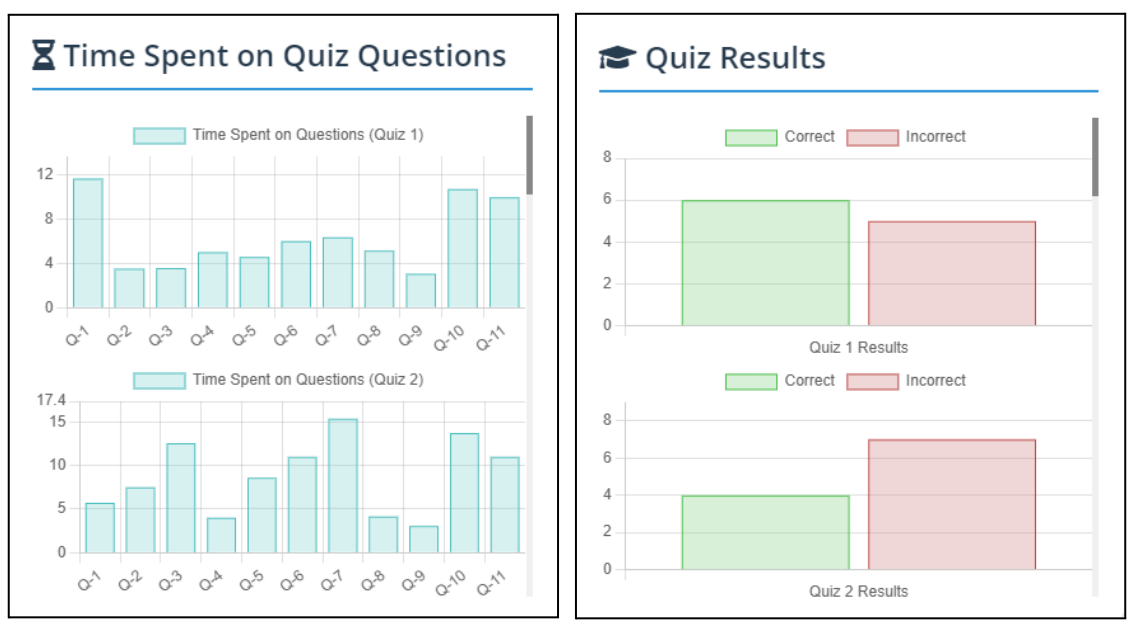

**Figure 5**. Screenshots from the instructor's dashboard (a) Time Spent on quiz questions and (b) quiz results of the respective quiz.

The bar graph (Figure 5a) shows time spent on each quiz question, highlighting question complexity and potential challenges. Figure 5b shows quiz outcomes, distinguishing correct and incorrect responses. These insights help educators identify areas needing more focus, supporting targeted teaching strategies and student improvement.

**Learning Progression and Bloom's Taxonomy**: The LA tool also incorporates a novel method of tracking students' learning progression by leveraging the few-shot learning capabilities of GPT-4. Each interaction a student has with the VirtualTA, including all messages and questions asked during a session, is analyzed through the lens of Bloom's Taxonomy - a widely accepted framework that categorizes learning objectives into different levels of complexity and specificity.

By using the GPT-4 model, we extract cognitive level indicators from student queries, allowing us to track changes in the depth and complexity of their understanding over time. For instance, questions indicative of recall or comprehension suggest an early stage of learning, while questions demonstrating application, analysis, synthesis, or evaluation suggest more advanced understanding. This progression mapping is based on digital planning verbs derived from bloom's taxonomy (Churches, 2010; Nikolić & Dabić, 2016). The digital planning verbs used for the learning progression are listed in Table A2. By integrating these verbs within the GPT-4 analytical framework, we can offer nuanced insights into the cognitive journey of the students. By observing these trends, educators can tailor their teaching strategies to better meet the changing needs of their students, offering a more personalized and effective learning experience.

**Figure 6**. Student query progression using bloom's taxonomy.

Figure 6 presents a detailed evolution on the progression and transformation of students' inquiries. Such a comprehensive understanding is invaluable, as it can facilitate the detection of learning trends, highlight topics that might necessitate additional clarification or emphasis, and consequently, guide the development of more effective, tailored teaching strategies. A disclaimer is included with the table, visible only when the table is hovered over, ensuring the layout remains clean and focused.

**Interactive Holistic Report**: As we transition into discussing the visualization of analytics data, it's worth noting the essential role that clear, impactful data visualization plays in the realm of LA. While collecting and analyzing comprehensive data is the foundation of any analytics tool (Sit et al., 2021; Xu et al., 2019), the true utility of such tools lies in their ability to translate this complex data into a form that is both digestible and actionable visual representations (Alabbad et al., 2022). With educators as the primary end-users of these insights, it becomes paramount to present the data in a manner that readily aids in decision-making, enhances understanding, and ultimately improves the learning experience for students. Presented in Figure 7 is the comprehensive instructor's dashboard, designed to provide a broad or focused view of student LA. Instructors can conveniently navigate through this interface, opting to delve into the learning patterns of a specific student or obtain an overarching understanding of the class's performance. This flexibility equips educators with a valuable tool to guide instructional decision-making based on individual or group LA.

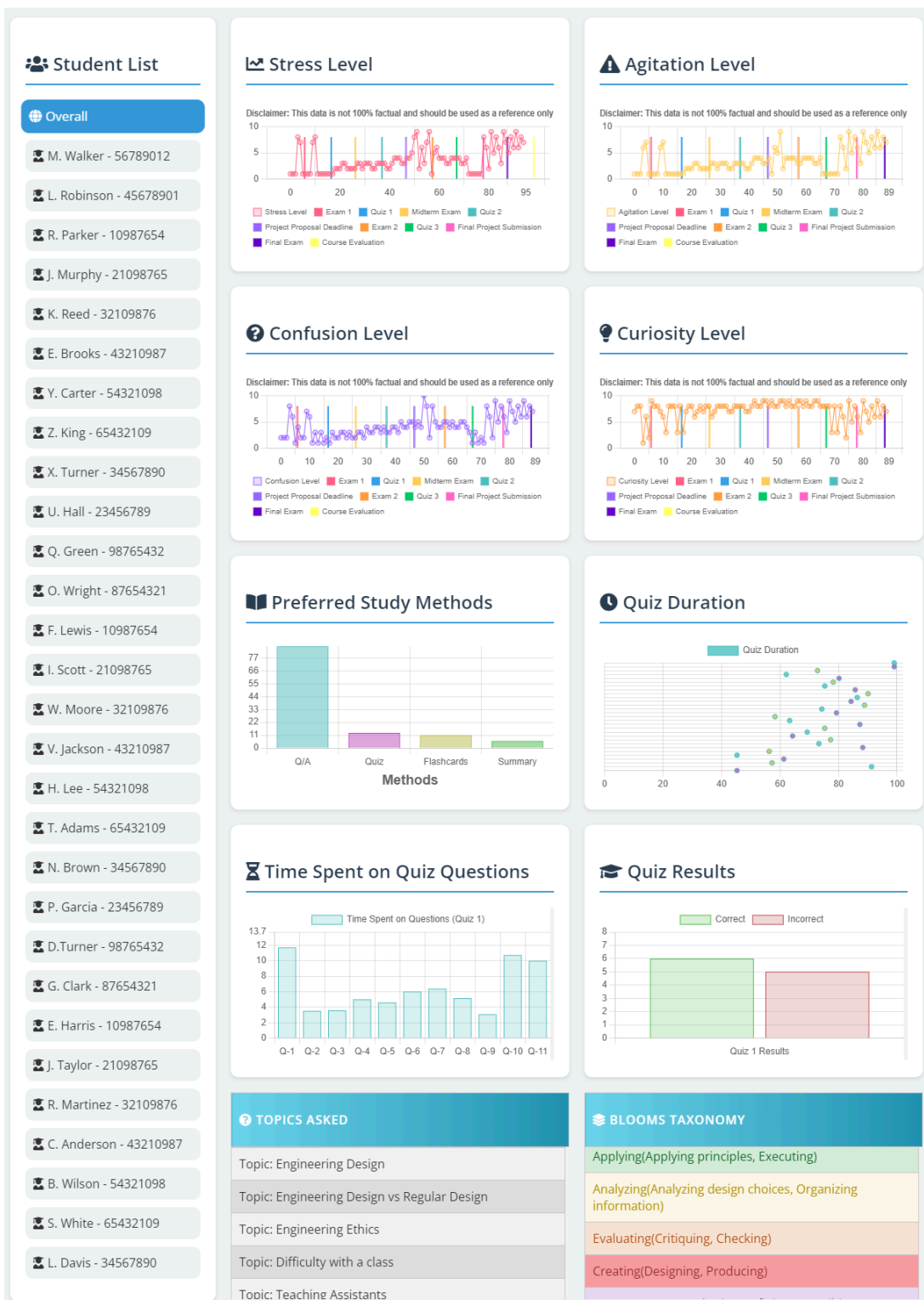

**Figure 7**. Instructor View of learning analytics dashboard – proof of concept with synthetic data

## 4.2.    Case Study Results: Faculty Feedback on the LA Tool

A case study was conducted with teaching faculty at the University of Iowa to gather feedback on the real-world applicability of the LA tool. An anonymous survey was administered to evaluate the tool's effectiveness, ease of integration, and concerns around privacy and data security. Since the survey questions were optional, there was a slight discrepancy in the number of responses for each question, with 1-2 responses varying.

As shown in Figure 8, respondents came from a diverse range of disciplines, with a significant proportion (38%) representing STEM fields such as Math, Science, and Engineering. This group's familiarity with technology likely influenced their feedback, emphasizing accuracy and reliability. Other respondents came from Professional Studies (14%), Humanities (11%), Social Sciences (8%), and Arts (2%), reflecting a range of pedagogical approaches. Notably, 27% of the respondents selected "Other," representing specialized fields like Nursing, Health Sciences, Medicine, Pharmacy, and Education, indicating that faculty in healthcare-related disciplines may have different expectations for AI tools due to the sensitive nature of their subject matter.

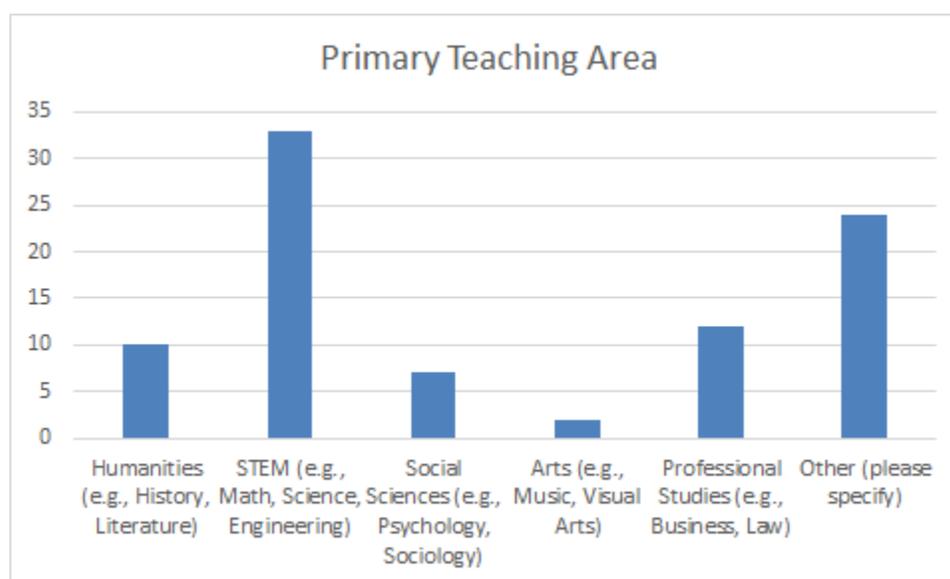

**Figure 8**: Primary Teaching Area of the Faculty (n=88)

As shown in Figure 9, regarding roles, Assistant Professors (30%), Professors (22%), and Associate Professors (22%) made up the majority of respondents, bringing seasoned perspectives based on long-term teaching experience. Graduate Assistants (7%) and Lecturers (3%) also contributed feedback, highlighting needs around ease of use and integration, while the 15% of the respondents selecting "Other" included adjunct instructors and postdoctoral researchers, adding further diversity to the perspectives on AI use in education.

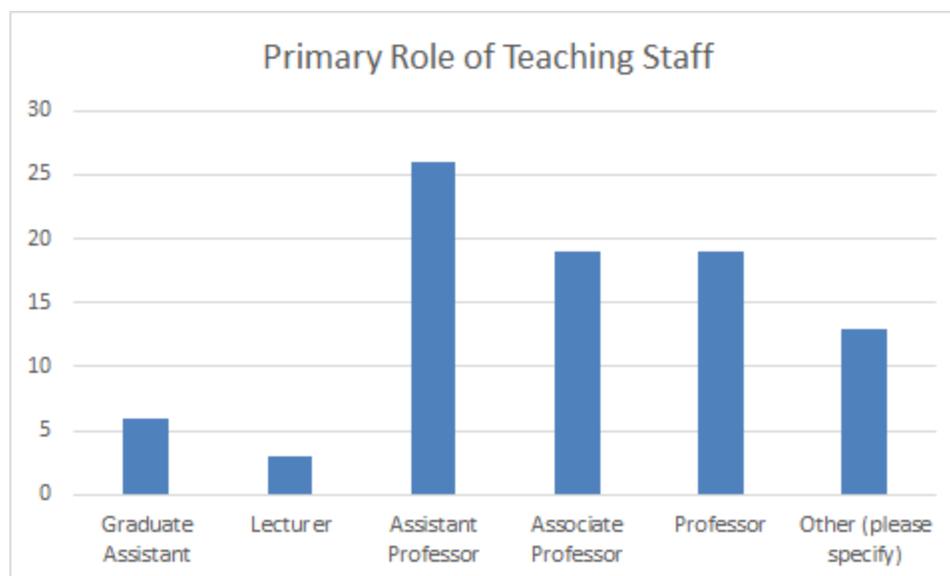

**Figure 9**: Primary Role of the Teaching Staff (n=86)

### 4.2.1. *Importance of Classroom Insights*

The data in Figure 10 highlights the key insights faculty find important for understanding student engagement and learning in the classroom. The highest priority for faculty was identifying frequently asked questions (80%) and tracking the progression of student learning over time (75%). These insights suggest that instructors value understanding both common areas of confusion and student learning trajectories, enabling them to tailor their teaching more effectively. Additionally, identifying common misconceptions (73%) was also a significant priority, further emphasizing the need for tools that help instructors address knowledge gaps.

Understanding students' preferred study methods (43%) also plays a role, though it is not as highly prioritized as other metrics. This suggests that while understanding how students approach their studies is important, it is secondary to more direct measures of learning progression and common misunderstandings. Less emphasis was placed on identifying student engagement with online tools (34%) and tracking timelines such as milestones (42%), indicating that while useful, these insights are not viewed as essential compared to other metrics.

Overall, the data shows that faculty prioritize insights that provide actionable information on student learning and areas of difficulty, which can directly inform their teaching practices.

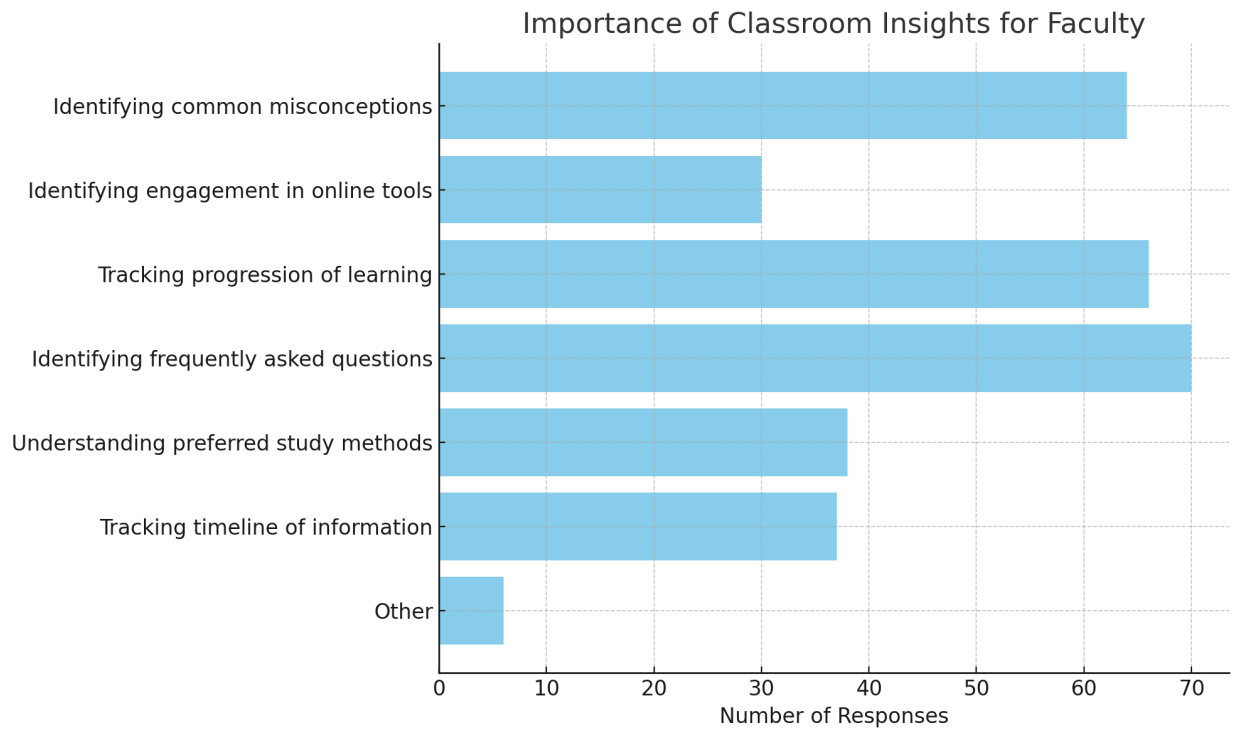

**Figure 10**: Importance of classroom insights (n=88)

### 4.2.2.　　　*Faculty Concerns: Accuracy of Insights vs. Privacy*

The combined data in Figure 11 illustrates faculty concerns regarding two critical factors when integrating AI tools into the classroom: student privacy and data security, as well as the accuracy of insights provided by these tools. Both issues are significant in determining the viability of AI-driven learning analytics.

A notable portion of respondents expressed either very concerned or extremely concerned responses about both privacy (51%) and accuracy (59%). This indicates that while AI tools hold potential for enhancing teaching, their adoption hinges on addressing these dual concerns. Faculty are especially cautious about the risks associated with handling sensitive student data and ensuring that the AI tools provide consistently accurate insights.

For privacy, it is clear that the majority of faculty see data protection as a paramount issue, with many concerned about how student data is stored, processed, and protected. Addressing this will require institutions to implement robust privacy safeguards and transparent data handling practices, ensuring that the faculty can trust the AI tools to manage student data responsibly. The minority who expressed less concern may be more familiar with the technology or trusting of current frameworks, but overall, privacy remains a crucial barrier to widespread AI adoption.

When it comes to the accuracy of insights, faculty appear just as cautious. Many are concerned about the reliability of AI-generated insights, reflecting a wariness about relying on these tools unless they can deliver consistently accurate and actionable data. Only a small fraction of faculty demonstrated low levels of concern, suggesting that a clear majority demand assurances about the validity of the insights before fully integrating AI tools into their teaching practices.

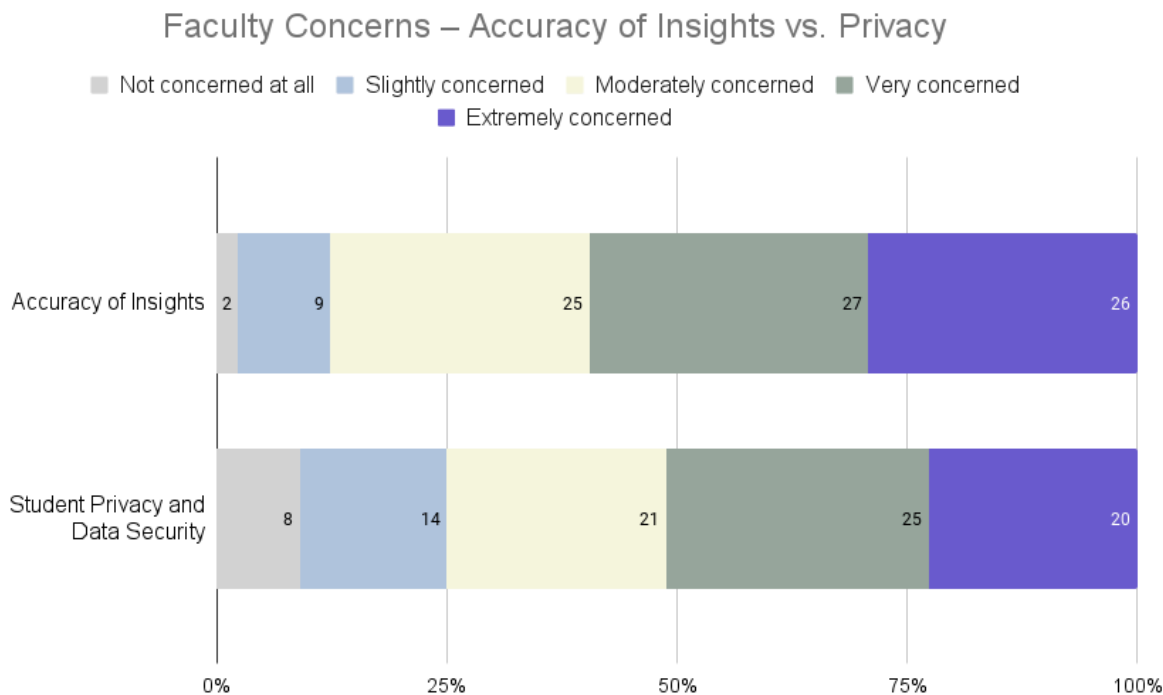

**Figure 11**: Faculty Concerns – Accuracy of Insights vs. Privacy (n=89 & n=88)

### *4.2.3. Additional Feedback and Suggestions*

Based on the additional feedback and suggestions provided by the participants, a clear theme emerges around the cautious yet optimistic stance towards the integration of AI in education. Many participants see potential benefits, particularly in saving time for students and aiding in deeper study processes. For example, AI could help students approach problems more thoughtfully, much like guidance from teaching assistants during office hours, rather than simply providing answers. This perspective highlights a desire to use AI in a way that enhances learning without replacing critical thinking.

However, concerns about the potential negative impacts of AI on student learning also surfaced. Some participants worry that over-reliance on AI could stifle critical thinking, emphasizing the need for students to engage in reasoning and problem-solving independently. Additionally, the fear that AI could be a barrier to developing skills in identifying misinformation was raised, suggesting a need for AI tools to focus on fostering students' ability to discern and reason with reliable information.

Data privacy and security were recurring concerns. Several participants questioned how AI handles sensitive student data, drawing attention to the necessity of transparent privacy policies and independent evaluations of AI tools. Some referenced recent high-profile privacy violations in technology sectors, underscoring the importance of ethical data management in educational tools.

The comments also indicate a practical concern: many educators feel pressed for time and would appreciate easier ways to incorporate AI into their teaching without needing to completely overhaul their courses. Additionally, some faculty are waiting for AI technology to mature, reflecting skepticism toward current systems like ChatGPT due to reliability issues.

## 4.3. Discussions

By comparing the results from both synthetic data and the feedback gathered from faculty, several key insights emerge. The synthetic data highlights the tool's technical capabilities, while faculty feedback emphasizes real-world applicability and concerns, particularly around data security and the accuracy of insights. One of the key features of the tool is its ability to measure both active and passive participation, providing nuanced insights into students' emotional states and learning progression. These capabilities enhance educators' ability to make informed, timely interventions, improving student engagement and learning outcomes. The integration of AI (GPT-4) further enriches this approach by allowing real-time sentiment analysis and topic identification, extending the tool's adaptability to various educational contexts and LMS platforms.

Both the synthetic data and faculty feedback underscored the value of real-time emotional analysis in improving student engagement. Faculty recognized the potential of such insights to tailor interventions and enhance classroom dynamics. While the tool's emotional analysis capabilities are promising, faculty feedback also highlighted that these features may not fully capture the complexities of human emotions and external factors affecting engagement. This suggests the need for further refinement to ensure the tool delivers more accurate and comprehensive emotional insights.

Faculty expressed strong concerns around the accuracy of AI-generated insights and the security of student data. These concerns align with the synthetic data results, which demonstrated the tool's ability to track learning patterns and engagement but also revealed limitations in consistency and reliability. Faculty feedback emphasized that accuracy and data security are critical factors for real-world applicability, particularly in fields like healthcare and STEM, where precision is essential. The real-world concerns highlight the need for more robust privacy measures, consistent accuracy, and transparency in how data is handled.

The tool's seamless integration with existing LMS platforms like Canvas was well received, but faculty emphasized the need for more intuitive data visualization to ensure ease of use. This feedback points to the necessity of refining the interface and reporting capabilities to better match the practical demands of educators, especially those who may be less familiar with data-driven tools. Additionally, the success of the tool depends on the digital literacy of both educators and students, underscoring the importance of comprehensive training to ensure effective use and interpretation of the generated data.

It is important to note that the dataset may be biased toward a more positive outlook on such technologies due to the overrepresentation of STEM faculty in the survey responses. STEM disciplines often have a higher familiarity with technological innovations, which may result in greater comfort and optimism about AI tools compared to fields like the arts, which were less represented in the sample. This skew in representation could mean that the perspectives of faculty in disciplines with less reliance on data-driven technologies, such as the humanities and arts, are underrepresented, potentially affecting the generalizability of the feedback.

A key limitation of this study is the absence of classroom experimentation. While the faculty survey provided valuable feedback on the potential applications of the LA tool, further evaluation is necessary in real classroom settings to fully understand its impact on student learning and engagement. Classroom experiments would offer critical insights into how effectively the tool integrates with teaching workflows, its practical benefits for students, and any unforeseen challenges that may arise during active use.

## 5.     Conclusion and Future Directions

This research successfully developed and deployed a novel learning analytics (LA) tool, representing a significant advancement in the field of educational technology. The tool, designed to integrate seamlessly with existing educational platforms like VirtualTA, captures, processes, and analyzes a wide array of data, offering educators a comprehensive understanding of students' engagement, performance, and learning patterns. By leveraging AI-driven capabilities, including real-time sentiment analysis and cognitive tracking based on Bloom's taxonomy, the tool enables a more data-driven and responsive approach to education. One of the strengths of the tool is its ability to measure both active and passive participation, providing nuanced insights that can lead to informed, timely interventions, ultimately improving student outcomes. However, while the tool demonstrates significant potential, further evaluations, particularly through classroom-based experimentation, are needed to validate its effectiveness in real-world teaching environments.

Moving forward, future research should focus on enhancing the tool's emotional analysis by integrating real-life classroom data and exploring attention-tracking algorithms. Classroom evaluation will be critical in determining how well the tool functions in diverse educational settings and with real students. Privacy and data security concerns must be prioritized through stronger anonymization techniques, transparent data usage policies, and adherence to data protection regulations. Additionally, the user interface and data visualization should be refined to offer more intuitive, actionable insights, facilitating easier use by educators. Expanding the tool's applicability to various educational levels, including special education, will further extend its versatility, allowing for broader adoption across different disciplines.

The continued exploration of emerging technologies, such as Virtual Reality, offers exciting opportunities to deepen the understanding of student behavior and engagement. Finally, developmental improvements in educator training and tool usability will be crucial for maximizing the tool's potential in delivering data-driven, personalized learning experiences. This research represents a foundational step in integrating AI into education and highlights the ongoing need for refinement to meet the practical, ethical, and technological demands of modern educational environments.

**Declaration of generative AI and AI-assisted technologies in the writing process**
During the preparation of this manuscript, the authors used ChatGPT, based on the GPT-4 model, to improve the flow of the text, correct grammatical errors, and enhance the clarity of the writing. The language model was not used to generate content, citations, or verify facts. After using this tool, the authors thoroughly reviewed and edited the content to ensure accuracy, validity, and originality, and take full responsibility for the final version of the manuscript.

**Appendix**

Table A1. Overview of Key metrics and indicators

| Metric | Description |
|---|---|
| **Engagement** | Engagement metrics encompass both active and passive aspects. Active engagement may be measured by the number of questions posed to the chatbot, the frequency of flashcard or quiz functionalities usage, and the total time spent interacting with learning materials. Passive engagement can be gauged by login frequency, session duration, and metrics as simple as page views (Yang, 2021). |
| **Stress** | Stress is generally inferred from certain proxy indicators, including the frequency of negative expressions and specific keywords hinting at factors such as increased workload, assignment anxiety, or concerns about exam performance in chatbot interactions (Thukral et al., 2020). It's not a direct measure but can offer valuable insights. |
| **Confusion** | Confusion can be identified by the frequency of repeated questions asked to the chatbot or the number of questions on a specific topic (Saqr et al., 2022), suggesting a difficulty in understanding that subject. |
| **Curiosity** | Curiosity could be measured by the variety of topics that a student inquiry about or the number of exploratory questions (IBM, 2021) they ask that extend beyond the prescribed syllabus. |
| **Learning Progression** | Utilizing Bloom's taxonomy (Bloom & Krathwohl, 1956), we assess the depth and sophistication of student queries, allowing us to discern shifts from foundational knowledge inquiries to analytical or evaluative questions. |
| **Agitation** | Identifying agitation involves tracking certain behaviors that could indicate heightened emotional states. Indicators might include the frequency of negative expressions and rapidly switching topics (IBM, 2021). These are not direct measures of agitation but can serve as proxies. |

Table A2. List of digital planning verbs

| Bloom's Categories | Verbs |
| --- | --- |
| **Remembering** | Copying, Defining, Finding, Locating, Quoting, Listening, Googling, Repeating, Retrieving, Outlining, Highlighting, Memorizing, Networking, Searching, Identifying, Selecting, Tabulating, Duplicating, Matching, Bookmarking, Bullet-pointing |
| **Understanding** | Annotating, Tweeting, Associating, Tagging, Summarizing, Relating, Categorizing, Paraphrasing, Predicting, Comparing, Contrasting, Commenting, Journaling, Interpreting, Grouping, Inferring, Estimating, Extending, Gathering, Exemplifying, Expressing |
| **Applying** | Acting out, Articulate, Reenact, Loading, Choosing, Determining, Displaying, Judging, Executing, Examining, Implementing, Sketching, Experimenting, Hacking, Interviewing, Painting, Preparing, Playing, Integrating, Presenting, Charting |
| **Analyzing** | Calculating, Categorizing, Breaking Down, Correlating, Deconstructing, Linking, Mashing, Mind-Mapping, Organizing, Appraising, Advertising, Dividing, Deducting, Distinguishing, Illustrating, Questioning, Structuring, Integrating, Attributing, Estimating, Explaining |
| **Evaluating** | Arguing, Validating, Testing, Scoring, Assessing, Criticizing, Commenting, Debating, Defending, Detecting, Experimenting, Grading, Hypothesizing, Measuring, Moderating, Probing, Predicting, Posting, Rating, Reflecting, Reviewing, Editorializing |
| **Creating** | Blogging, Building, Animating, Adapting, Collaborating, Composing, Directing, Devising, Podcasting, Wiki Building, Writing, Filming, Programming, Simulating, Role Playing, Solving, Mixing, Facilitating, Managing, Negotiating, Leading |